\documentclass[sn-mathphys-num]{sn-jnl}

\usepackage{graphicx}%
\usepackage{multirow}%
\usepackage{amsmath,amssymb,amsfonts}%
\usepackage{amsthm}%
\usepackage{mathrsfs}%
\usepackage[title]{appendix}%
\usepackage{xcolor}%
\usepackage{textcomp}%
\usepackage{manyfoot}%
\usepackage{booktabs}%
\usepackage{algorithm}%
\usepackage{algorithmicx}%
\usepackage{algpseudocode}%
\usepackage{listings}%
\usepackage{verbatim}
\usepackage{enumitem}
 \usepackage{comment}

\theoremstyle{thmstyleone}%
\theoremstyle{thmstyletwo}%

\theoremstyle{thmstylethree}%

\begin{document}

\title[Article Title]{The Impact of Social Attractiveness on Casual Group Formation: Power-Law Group Sizes and Suppressed Percolation}

\author[1]{\fnm{Matheus S. } \sur{Mariano}}

\author[1]{\fnm{Jos\'e F.} \sur{Fontanari}}

\affil[1]{\orgdiv{Instituto de F\'{\i}sica de S\~ao Carlos}, \orgname{Universidade de S\~ao Paulo}, \orgaddress{\city{S\~ao Carlos}, \postcode{13560-970}, \state{S\~ao Paulo}, \country{Brazil}}}

\abstract{The dynamics of casual group formation has long been a subject of interest in social sciences. While early stochastic models offered foundational insights into group size distributions, they often simplified individual behaviors and lacked mechanisms for heterogeneous social appeal. Here, we re-examine the attractiveness-driven interaction model, an agent-based framework where point-like agents move randomly in a 2D arena and exhibit varied social appeal, leading them to pause near highly attractive celebrity peers. We compare this model to a null model where the agents are continuously in movement, which resembles a Random Geometric Graph. Our extensive simulations reveal significant structural and dynamic differences: unlike the null model, the attractiveness-driven model's average degree increases linearly with system size for fixed density, resulting in more compact groups and the suppression of a percolation transition. Crucially, while the null model's group size distribution is either exponentially decaying or bimodal, the attractiveness-driven model robustly exhibits a power-law distribution,  $P(n) \propto n^{-2.5}$, with an exponent independent of density. These findings, obtained through computationally intensive simulations due to long equilibration times, offer a thorough quantitative characterization of this model, highlighting the critical role of individual attractiveness in shaping social aggregation in physical space.
}


\maketitle

\section{Introduction}\label{sec:intro}

Human social life in public settings is fundamentally characterized by the dynamic formation and dissolution of temporary,  face-to-face interactions,  which spontaneously give rise to casual groups \cite{Coleman_1964, Burgess_1984}.  The intricate ebb and flow of individuals forming and leaving these free-forming clusters---whether observed in a bustling cocktail party or the steady stream of pedestrians on a sidewalk---has long captivated researchers across social sciences \cite{Coleman_1961,Cohen_1971}.  Pioneering stochastic models from the 1960s laid crucial groundwork for understanding the size distributions of such groups,  often by simplifying individual behaviors and abstracting away specific personal attributes \cite{White_1962}.  However, the recent advent of large-scale empirical data \cite{Cattuto_2010}  and sophisticated agent-based modeling techniques  has opened unprecedented avenues for a more nuanced and quantitative understanding of these complex social systems.  Our work bridges these classical sociological insights with contemporary computational approaches  by characterizing the statistical properties of casual groups within a model that explicitly incorporates heterogeneous social appeal among agents  \cite{Starnini_2013,Starnini_2016},   a critical factor largely absent from earlier theoretical constructs \cite{Coleman_1961,Cohen_1971,White_1962}.

Specifically,    we re-examine an agent-based model  for  face-to-face interactions developed by Starnini et al.  \cite{Starnini_2013, Starnini_2016},  which posits point-like agents moving randomly within a two-dimensional square arena.  A key feature of this model is the heterogeneous attractiveness assigned to agents, where some exhibit particularly high appeal,  who we refer to as celebrities.   Agents are more likely to pause their random movement and thus prolong interaction when they are within the interaction zone of these attractive peers.  It is due to this core interaction mechanism that we refer to the model as the attractiveness-driven interaction model.   While the original studies primarily focused on the temporal properties of face-to-face interactions, such as contact durations and inter-event times,  our research shifts the emphasis to the equilibrium properties of the emergent spatial contact network,  where groups are naturally defined as connected components within this graph \cite{Newman_2018}. 

To fully appreciate the impact of celebrity presence,  we compare the attractiveness-driven interaction model against a null model.  In this null model,  agents are continuously in motion,  rendering agents'  attractiveness,  whether heterogeneous or not,   irrelevant to interaction durations.  This null model shares similarities with the classic Random Geometric Graph (RGG) \cite{Gilbert_1961,Penrose_2003}.  Our findings reveal a stark contrast: while the average degree of the null model depends solely on density,  the average degree in the attractiveness-driven interaction model increases linearly with the arena's linear size for a fixed density.  This implies that as the system scales up,  the clusters forming around celebrity agents become increasingly compact.  Crucially,  this enhanced clustering around celebrities actively prevents the formation of a single,  large cluster encompassing most agents,  thereby eliminating the percolation transition that characterizes  the  fraction of agents in the largest group in both the null model and the RGG  \cite{Dall_2002}.

Beyond these network-level properties,  a particularly important result concerns the distribution of group sizes $P(n)$, which is widely considered the preferred empirical measure for characterizing casual groups \cite{Coleman_1961,Cohen_1971,White_1962}.  Our analysis reveals a distinct behavior across the models: while the null model yields a group size distribution that decays exponentially with increasing group size $n$ for  low densities, transitioning to a bimodal distribution at higher densities \cite{Dall_2002},  the attractiveness-driven interaction model surprisingly produces a power-law distribution,  $P(n) \propto n^{-\beta}$ with $\beta = 2.5$.   Notably,  the power-law exponent $\beta$ appears to be independent of density,  highlighting a robust emergent property driven by heterogeneous social appeal.

While our study focuses on casual groups, it is useful to distinguish them from the more stable concept of communities prevalent in complex network analysis.  Within the framework of complex networks, a community is typically defined as a densely interconnected group of nodes that has sparser connections to the rest of the network \cite{Newman_2018}.  These communities are ubiquitous in social and biological systems and are often understood to represent relatively enduring social groupings,  driven by shared interests,  background,  or persistent relationships \cite{Girvan_2002}.  However, a crucial distinction from casual groups lies in their temporal stability: complex network communities are generally characterized by their relative persistence over time. This inherent stability makes them fundamentally different from the transient,  fleeting nature of casual groups,  which are continuously forming and dissolving.  Consequently, despite their utility in characterizing more permanent social structures,  such communities are not ideal models for the rapidly evolving dynamics of casual groups, which are better captured by models of face-to-face interaction networks.

The remainder of this paper is organized as follows. In section \ref{sec:model}, we provide a detailed description of the asynchronous agent-based model dynamics, including the rules governing agent movement and interaction. Section \ref{sec:res} presents our comprehensive simulation findings. Particular attention is drawn to the exceptionally large timescales required for the attractiveness-driven interaction model to reach equilibrium---a state where macroscopic quantities such as the average degree and the fraction of agents in the largest group stabilize around time-independent values. Consequently, studying the attractiveness-driven interaction model proves to be highly computationally intensive, even for relatively small arena sizes. Finally, section \ref{sec:disc} summarizes our main findings and presents future work directions.

\section{The model}\label{sec:model}

Following Starnini et al.  \cite{Starnini_2013,Starnini_2016},  we investigate the  properties of groups emerging from  face-to-face interactions within  a system of $N$ point-like agents.  These agents  move in a square arena of linear size $L$,  with periodic boundary conditions,  resulting in a density $\rho=N/L^2$.  Interactions occur between agents within a distance  $d$,   and agents  perform a random walk,  moving a fixed step of length $v$ in a randomly chosen direction.  Consistent with the original model,  we set $v=d=1$.  

A crucial feature of this model is that each agent $i \in \{1, \ldots,N\}$  possesses an intrinsic attractiveness or social appeal $a_i \in [0,1]$. This value quantifies an agent's ability to capture the interest of others, specifically by influencing their willingness to cease movement and maintain an interaction. The model also distinguishes between two classes of agents: active and inactive.  Inactive agents become active with a probability $r_i \in [0,1]$.  Only active agents can interact and move; they transition to an inactive state with probability $1-r_i$ when isolated.    Both the attractiveness $a_i$ and the activation probability $r_i$ are assigned randomly from the unit interval at time $t=0$  for each agent and remain fixed throughout the simulation.

The dynamics of this asynchronous face-to-face interaction model begins with agents randomly distributed in the arena at $t=0$ and proceed as follows.
At each time step $ \delta t = 1/N$,  a focal agent,  say agent $i$,  is randomly selected.
\begin{itemize}
\item If  agent $i$ is inactive:  it  attempts to become active with probability $r_i$.   If this attempt fails,   it remains inactive.  
\item If  agent $i$ is active: 
\begin{itemize}
\item Decision to move: agent $i$ assesses the attractiveness of all other active agents within its interaction distance $d$.   It then decides to move with a probability $p_i(t)$ defined as
\begin{equation}\label{pit}
p_i(t) = 1 -  \max_{j \in \mathcal{N}_i(t)}  \left \{ a_j \right \} ,
\end{equation}
where $\mathcal{N}_i(t) $ represents the set of active neighbors of agent $i$ at time $t$.  Notably,  an active agent with no active neighbors (i.e.,  $\mathcal{N}_i(t) = \emptyset$) will always choose to move. 

\item Movement: if agent $ i$  decides to move,  it selects a random direction within $[0,2\pi)$ and moves a step of length $v$.

\item Inactivation: after moving,  if agent $i$ finds itself isolated (i.e.,  no other active agents are in its neighborhood),  it may become inactive with probability $1-r_i$.
\end{itemize}
\end{itemize}

The asynchronous update scheme,  characterized by a time increment $ \delta t = 1/N$,  ensures that exactly $N$ agents are selected as focal agents during the interval from $ t$  to $ t+1$,  though some agents may be selected multiple times.

The structure of the groups formed by these face-to-face interactions is illuminated by creating an edge between any two active agents within distance $d$,   thereby forming an undirected contact graph.  A group is defined as a connected component within this graph \cite{Newman_2018},  meaning it is a subgraph where every active agent can reach every other active agent within that subgraph.  Inactive agents do not interact,  are invisible to other agents,  and thus do not belong to any group.  They can be conceptualized as agents momentarily outside the interaction arena.

Our definition of a casual group is based on the connected components of the instantaneous interaction graph. This aligns with the fundamental interpretation of a spatial aggregation of agents interacting via proximity, which forms the basis for the group size analysis in the original attractiveness-driven interaction model by Starnini et al. \cite{Starnini_2013,Starnini_2016}.  While we acknowledge that more restrictive definitions, such as maximal cliques  \cite{Cencetti_2021} or density-based definitions  \cite{Menardi_2022}, are used in other contexts, our choice is motivated by its relevance to studying percolation and the global clustering effects inherent to this specific mobility model.

The most distinctive and powerful feature of the agent-based model proposed by Starnini et al.  \cite{Starnini_2013,Starnini_2016}  lies in its incorporation of heterogeneous social appeal among agents. This heterogeneity allows us to conceptually frame agents with significant social appeal as celebrities,  providing a natural and elegant generalization of the casual group models developed in the 1960s \cite{Coleman_1961, White_1962} (for a more recent contribution on casual groups, see \cite{Fontanari_2023,Fontanari_2024}).  Unlike those earlier models,  which typically described temporary groups formed in settings not dominated by a few highly influential individuals \cite{White_1962},  the attractiveness-driven interaction model explicitly accounts for such celebrity effects.

\section{Results}\label{sec:res}
To thoroughly understand the impact of celebrity presence on temporary group statistics,  our findings are compared against a null model.  In this null model,  we set  $a_i=0$ for all agents,  meaning active agents never cease movement. This setup bears strong resemblances to,  though is not identical to,  the classic Random Geometric Graph (RGG). Introduced in the 1960s to model wireless communication networks \cite{Gilbert_1961,Penrose_2003},  the RGG's properties,  particularly its threshold phenomena,  have been extensively explored in statistical physics  \cite{Dall_2002}.  Furthermore,  the RGG has become a preferred framework for modeling various phenomena, including  synchronization \cite {Diaz_2009},  opinion dynamics \cite{Zhang_2014, Reia_2020},  epidemic spreading \cite{Estrada_2016},   and cooperative problem-solving  \cite{Gomes_2019}  in scenarios where agents move randomly within geographically constrained regions.  A key characteristic of the RGG is its spatially homogeneous distribution of agents,  uniformly and randomly placed within the square arena,  which facilitates the derivation of  analytical results \cite{Penrose_2003}. The primary distinction between our null model and the RGG lies in the treatment of isolated agents: the RGG typically features more isolated agents,  since in our null model,  isolated agents can become inactive and effectively be removed from the contact graph.  Nevertheless,  at sufficiently high densities where isolated agents are rare,  both models yield identical results.

\begin{figure} 
\centering  
 \includegraphics[width=0.8\textwidth]{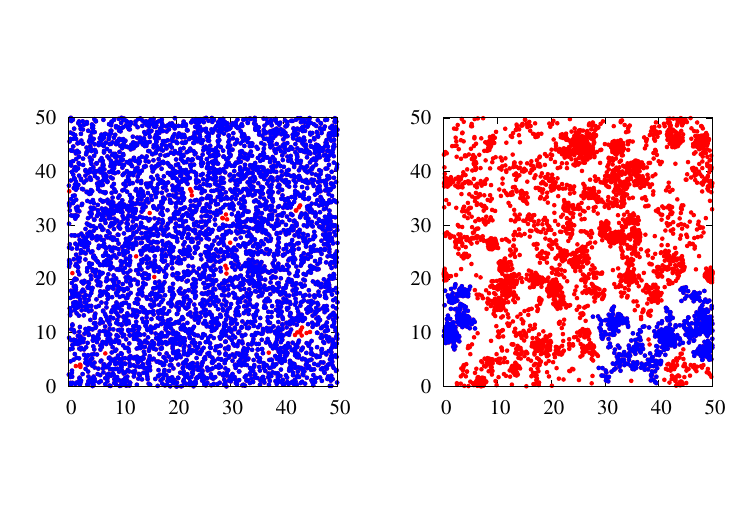}  
 \caption{Spatial distribution of active agents at $t=10^5$.  Snapshots illustrate the spatial distribution of active agents (red symbols) for the null model (left panel) and the attractiveness-driven interaction model (right panel).  Active agents belonging to the largest group are highlighted in blue.  In the null model,  the largest cluster contains $4890$ agents,  compared to $1226$ agents in the attractiveness-driven interaction model.   Simulation parameters are $L=50$ and $N=5000$. }  
\label{fig:1}  
\end{figure}
\begin{figure} 
\centering  
 \includegraphics[width=0.8\textwidth]{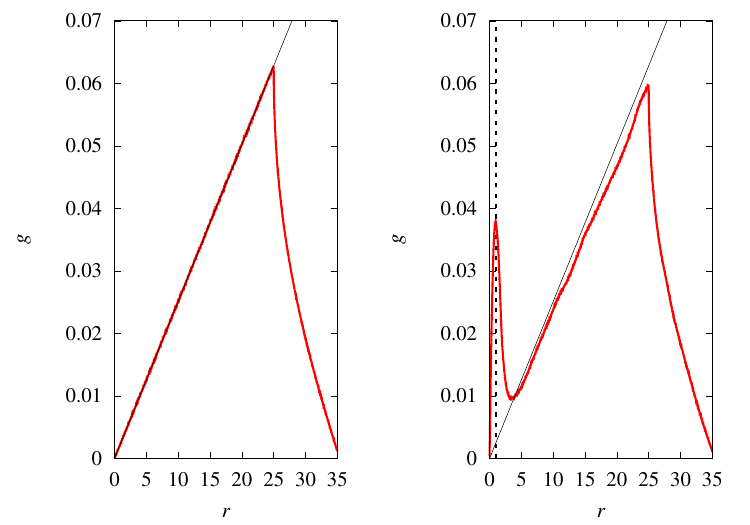}  
 \caption{Pair correlation function $g (r)$ at  $t=10^5$.  The pair correlation function is shown for the null model (left panel) and the attractiveness-driven interaction model (right panel).  Results are averaged over 100 independent simulation runs. The thin black line represents the analytical result for uniformly randomly distributed agents,  $g (r) = 2\pi r/L^2$,  which is valid  for $ r <L/2$.  The vertical dashed line indicates $r=d=1$.  Simulation parameters are $L=50$ and $N=5000$.  }  
\label{fig:2}  
\end{figure}
 Figure \ref{fig:1} presents snapshots of the spatial distribution of active agents at $t=10^5$  for the null model and the attractiveness-driven interaction model,  at a density $\rho = 2$.   At this density,  the null model typically shows nearly all agents belonging to a single large group.  In stark contrast,  the attractiveness-driven interaction model reveals agents dispersed into several groups of comparable sizes.  To quantitatively assess this inhomogeneity in spatial distribution,   we calculate the pair correlation function $g (r)$,  which represents the fraction of pairs of active agents whose distance falls between  $r$ and $r + \delta r$. 
Due to periodic boundary conditions,  the maximum possible distance between any two agents is $L/ \sqrt{2}$.  
Importantly,  $g(r)$ is independent of $N$.  In fact,   $N$ solely determines the number of samples,  $N(N-1)/2$,  used to estimate  $g(r)$ in each simulation run.  For uniformly randomly distributed agents,   $g (r) = 2\pi r/L^2$  for $ r < L/2$,  a range where boundary conditions do not affect distance calculations.   Figure \ref{fig:2}  displays the pair correlation function   computed using $\delta r = 0.01$.  The results for the null model  closely align with the pair correlation function of uniformly randomly distributed agents for  $r < L/2$.  Conversely,  the attractiveness-driven interaction model exhibits a pronounced peak at $r=d=1$ in its  pair correlation function,  quantitatively confirming the significantly higher occurrence of closely clustered agents,  as qualitatively observed in Fig.\  \ref{fig:1}.  The dependence of this clustering effect on the agent density is further detailed in the Supplementary Material (Fig.\  \ref{fig:S4}).

\begin{figure} 
\centering  
 \includegraphics[width=0.8\textwidth]{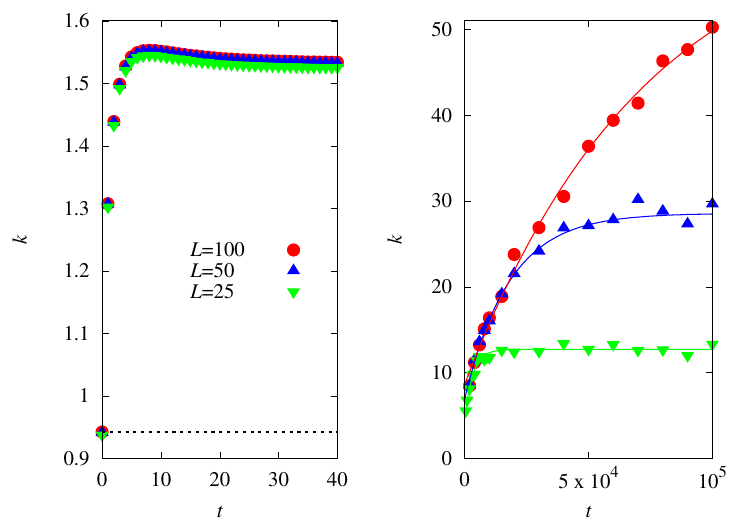}  
 \caption{Time evolution of average degree.   Average degree $k$ as a function of time $t$  for the null model (left panel) and the attractiveness-driven interaction model (right panel). Results are shown for $\rho=0.6$ and varying linear sizes  $L=25,  50$,  and $100$.  Data points (symbols) represent averages obtained from $10^3$   independent simulation runs.  In the left panel,  the horizontal dashed line indicates the analytical prediction $k = \pi \rho/2$.  In the right panel,  the solid colored curves correspond to the exponential fit function given by eq.\ (\ref{fitk}).  }  
\label{fig:3}  
\end{figure}

A significant aspect of studying the attractiveness-driven interaction model is the time required for the dynamics to reach a stationary regime.  In this regime,  macroscopic parameters, such as the graph's average degree $k$,  exhibit only small fluctuations around a stable,  time-independent value.  Figure \ref{fig:3} illustrates the time dependence of $k$ at a fixed density of $\rho=0.6$. The left panel,  presenting results for the null model,  shows that the system quickly attains a stationary state,  which is largely independent of the arena's linear size $L$ as long as $L$  is not too small.  These results highlight a key difference between our null model and the RGG.  In the RGG,  the $N$  agents are assumed to be homogeneously distributed in the arena,  leading to an average degree equal to the mean number of agents within a unit radius circle centered on a focal agent, i.e., $k = \pi \rho$. Considering that in both our null and attractiveness-driven interaction models agents are initially randomly distributed at $t=0$,  but,  on average,  only half are active,  our initial prediction would be $k = \pi \rho/2$.  This prediction perfectly aligns with our simulation results at $t=0$.  However,  as the agents move randomly within the arena,  the mean degree of the null model becomes considerably greater than this initial prediction from the homogeneous distribution assumption.  This deviation occurs because only isolated agents can become inactive,  and  since our undirected graphs are formed exclusively by active agents,  there are effectively fewer isolated agents in our null model compared to a pure RGG, consequently leading to a higher average degree.

In contrast to the null model,  the attractiveness-driven interaction model necessitates considerably longer timescales to achieve equilibrium.  Both the equilibrium values and the time required to reach this state exhibit a strong dependence on $L$,  as depicted in the right panel of Fig.\  \ref{fig:3}.  These quantities can be estimated by fitting the time evolution of the average degree $k(t)$ to the exponential function
\begin{equation}\label{fitk}
k(t) = k_\infty -  k^*  \exp \left (- t/\tau_k \right  )
\end{equation}
in the region $t \in [ 500,2\times10^{5}]$.   Here,   $k_\infty$,  $k^*$,    and $\tau_k$  are the fit parameters that vary with the density $\rho$  and the  linear size $L$.    
We observe that the characteristic relaxation time $\tau_k$ increases proportionally with $L^2$  (or with $N$,  given a fixed density).  Evidence for the $L^2$ scaling is presented in Fig.\ \ref{fig:S1} of the Supplementary Material. This significant dependence severely constrains the maximum system size that can be feasibly studied through simulations.  This behavior reflects the inherent difficulty in breaking down an initially homogeneous configuration of agents into distinct clusters centered around a few celebrities,  as well as the ongoing competition among these influential agents for followers.

\begin{figure} 
\centering  
 \includegraphics[width=0.8\textwidth]{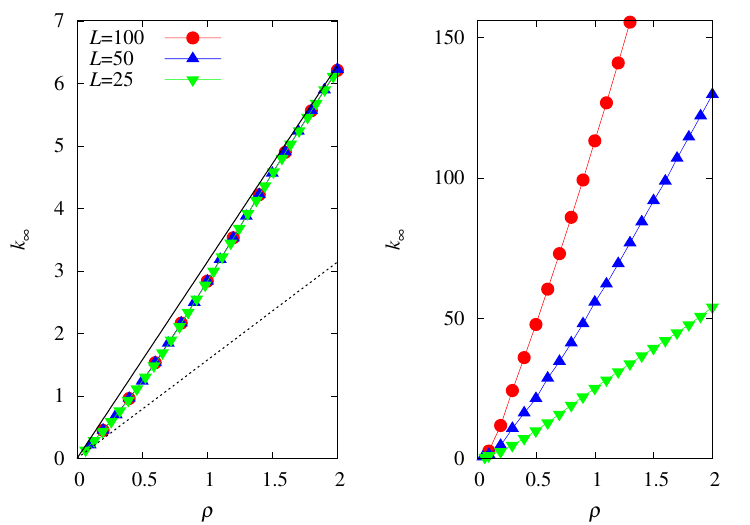}  
 \caption{Equilibrium average degree as a function of density.  Equilibrium average degree $k_\infty$  for the null model (left panel) and the attractiveness-driven interaction model (right panel) as functions of density $\rho$.  Results are shown for linear sizes  $L=25,  50$,  and $100$.  Data points (symbols) represent averages obtained from $10^3$
  independent simulation runs.   In the left panel,   the solid black line denotes $k = \pi \rho$,  while the dashed black line indicates $k = \pi \rho/2$.  Lines connecting the symbols serve as visual guides. }  
\label{fig:4}  
\end{figure}

Figure \ref{fig:4} illustrates the dependence of the equilibrium average degree  $k_\infty$  on the density $\rho$. The left panel demonstrates that for the null model $k_\infty$ depends solely on density,  consistent with the observations in Fig.\  \ref{fig:3}.  However,  its specific dependence on $\rho$ diverges from that of a pure RGG.  Specifically,  for an RGG, the average degree is $k = \pi \rho$ if all agents are active,  and $k = \pi \rho/2$ if,  on average,  only half of the agents are active.  Our null model aligns with RGG predictions at density extremes: at high densities,  where isolated agents are rare,  most agents remain active,  leading to results consistent with $k = \pi \rho$.  Conversely,  at low densities,  where many agents are isolated and thus often inactive,  results align with $k = \pi \rho/2$.

\begin{figure} 
\centering  
 \includegraphics[width=0.8\textwidth]{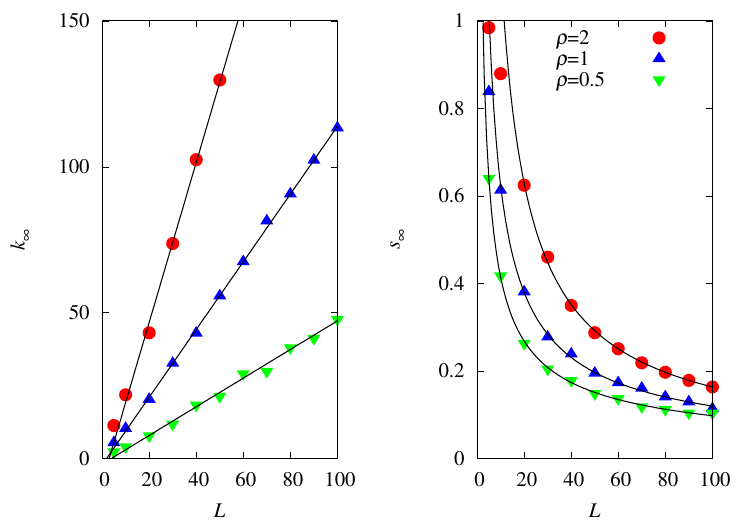}  
 \caption{Equilibrium average degree and largest group fraction vs.  linear system size. 
 Equilibrium average degree   $k_\infty$ (left panel) and fraction of active agents in the largest group $s_\infty$ (right panel)     for  the attractiveness-driven interaction model as  functions of the linear  size $L$ of the square arena.   Results are shown for densities    $\rho=0.5,  1$ and $2$.  Data points (symbols) represent averages obtained from $10^3$  independent simulation runs.    In the left panel,   the  straight lines depict the linear fit $k_\infty =  aL + b$.  In the right panel,  the curves illustrate the power-law fit $s_\infty =  aL^{-b}$,  where $a$ and $b$ are fit parameters for both cases.  The fitting region for all curves is $L > 20$.}  
\label{fig:6}  
\end{figure}

The right panel of Fig.\  \ref{fig:4} presents the results for the attractiveness-driven interaction model.  For a fixed $L$,   increasing $\rho$ is equivalent to increasing $N$,  so  a linear increase in $k_\infty$ with respect to $\rho$ is generally expected,  provided the density is not too low.  As noted previously,  at very small densities,  an increase in $N$ does not directly translate into a proportional increase in the number of active nodes in the graph due to the potential for isolated agents to become inactive.  The  remarkable finding, however,  is the increase in $k_\infty$ with increasing $L$ when the density $\rho$ is held constant  (as clearly shown in Fig.\  \ref{fig:6}). This indicates that as  the number of agents $N$ and the  linear size $L$ increase simultaneously  while maintaining a constant density,  the groups forming around celebrities become increasingly compact.
While the resulting network topology is characterized by an average degree $k_\infty$,  a more detailed view of the instantaneous network structure is given by the full degree distribution,  $P(k)$, which we show in the Supplementary Material (Fig.\  \ref{fig:S5}) to be a single-peaked, broad distribution that decays exponentially for large $k$.

\begin{figure} 
\centering  
 \includegraphics[width=0.8\textwidth]{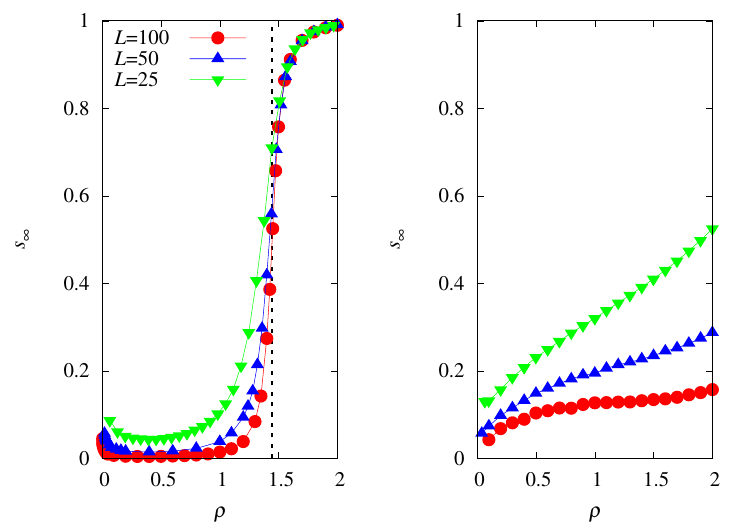}  
 \caption{Equilibrium fraction of active agents in the largest group.   Equilibrium fraction of active agents in the largest group $s_\infty$  for the null model (left panel) and the attractiveness-driven interaction model (right panel) as function of density $\rho$.  Results are shown for linear sizes $L=25,  50$,  and $100$.  Data points (symbols) represent averages obtained from $10^3$ independent simulation runs.  In the left panel,  the vertical dashed line indicates the critical density $\rho_c \approx 1.44$ of the Random Geometric Graph (RGG).  Lines connecting the symbols serve as visual guides. }  
\label{fig:5}  
\end{figure}

Another quantity of significant interest,  particularly from the perspective of statistical physics  \cite{Stauffer_1994},  is the fraction of active agents belonging to the largest group.  We denote this fraction by $s$,  which ranges from $1/N_{act}$ (indicating all groups are isolates) to $1$ (implying all active agents belong to a single group).  Here,  $N_{act}$
 represents the total number of active agents,  a value dependent on both $\rho$ and $L$.  The snapshots presented in Fig.\  \ref{fig:1} qualitatively illustrate the striking difference in the largest group size between the null and attractiveness-driven interaction model at high density.  To quantitatively estimate the equilibrium value of $s$,  denoted as $s_\infty$,  we employ a procedure analogous to that used for $k_\infty$.  Specifically,  we measure $s$ at various times and fit the resulting data to the exponential function
\begin{equation}\label{fits}
s(t) = s_\infty -  s^*  \exp \left (- t/\tau_s \right  )
\end{equation}
for values of $t$ in the region $t \in [ 500,2\times10^{5}]$.    As with the previous fit,   $s_\infty$,  $s^*$,   and $\tau_s$  are the extracted fit parameters.   The fitting interval for both  $k(t)$  and $s(t)$ starts after an initial transient period (e.g.,  $t > 500$) to exclude the early non-monotonic behavior, which reflects the system's necessary self-organization from the homogeneous initial conditions (RGG-like configuration) to the final heterogeneous, attractiveness-driven equilibrium.

The left panel of Fig.\  \ref{fig:5} presents the results for the null model.   In this model,   $s_\infty$ exhibits a discontinuity at a critical density $\rho_c$
in the thermodynamic limit.  Specifically,  as $L \to \infty$,  $s_\infty \to 0 $  for $\rho < \rho_c$,   and  $s_\infty $ becomes positive otherwise.  For high densities, where the null model is effectively identical to the RGG,  the figure shows an estimated  $\rho_c \approx 1.44$ 
for the RGG \cite{Dall_2002}.   Interestingly,  at very low densities,  $s_\infty$  is observed to increase as $\rho$ decreases for a fixed $L$.  This counter-intuitive behavior occurs because the number of active agents,  $N_{act}$ (which forms the denominator in the definition of $s$),  decreases significantly at low densities due to a higher proportion of  agents becoming inactive.  Thus,   it is not that the number of active agents within the largest group increases.  Rather,  the total pool of active agents shrinks,  leading to a larger fraction of these agents belonging to the dominant group.

In contrast,  the results for the attractiveness-driven interaction model,  shown in the right panel of Fig.\  \ref{fig:5},  do not exhibit a threshold phenomenon.  However,  they strongly support the scenario of highly compact clusters of followers forming around a few celebrities.  The slow and smooth increase in $s_\infty$ with $\rho$  indicates that active agents are distributed relatively uniformly among these compact groups without significantly increasing their spatial extension.  Such an uncontrolled spatial increase would eventually lead to a sharp percolation transition similar to those observed in the null model and the RGG.  While groups will eventually begin to merge at sufficiently high densities, causing $s_\infty$ to approach $1$,  this process is notably slow and smooth,  distinctly lacking the threshold  behavior of the null model.

Figure \ref{fig:6} summarizes the dependence of the equilibrium quantities $k_\infty$ and $s_\infty$ on the linear size $L$ of the square arena for the attractiveness-driven interaction model.  As shown in the left panel,  the average degree $k_\infty$ increases linearly with $L$ (or proportionally to $N^{1/2} $  when density $\rho$ is fixed).  This implies that,  on average,  the number of active agents within an interaction distance $d=1$ from an active agent scales with $N^{1/2}$.
The right panel reveals that $s_\infty$  decays as a power law,  specifically as $1/L^\gamma$,  when $L$ increases. The exponent $\gamma$  ranges from approximately $0.6$ for $\rho=0.5$  to $0.8$ for $\rho=2$.  Precisely determining this power-law decay and accurately estimating the exponent $\gamma$ would,  of course,  necessitate simulations with significantly larger values of $L$ than those considered here.  Assuming that the number of active agents $N_{act}$ grows linearly with $N$ (see Fig.\ \ref{fig:S2} of the Supplementary Material),  the absolute number of agents in the largest group  (i.e.,  $s_\infty N_{act} $) scales approximately with $N^{1- \gamma/2}$ for fixed agent density.

\begin{figure} 
\centering  
 \includegraphics[width=0.8\textwidth]{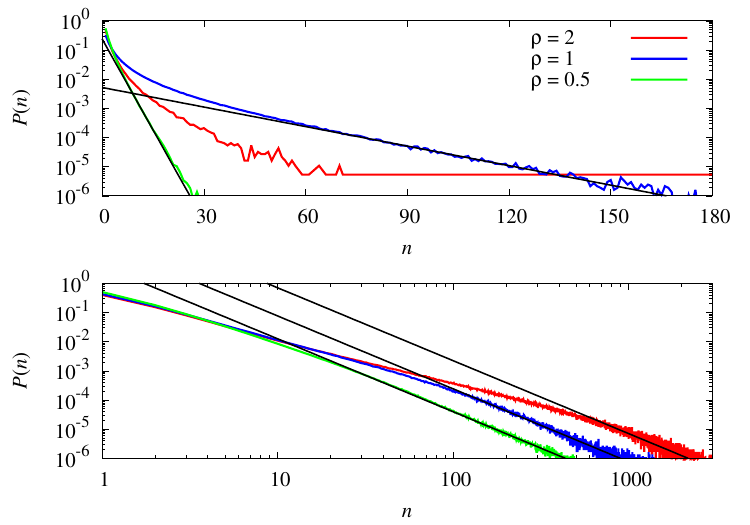}  
 \caption{Group size distribution at $t=10^5$.  Distribution of group sizes $P(n)$  for the null model (top panel) and the attractiveness-driven interaction model (bottom panel) with linear size $L=50$.  Results are shown for densities $\rho = 0.5, 1$,  and $2$.  These  distributions were estimated using $10^4$ to $10^5$  independent simulation  runs.   In the top panel,  the black lines represent exponential fits to the data for $\rho = 0.5$ and $1$.   
 In the bottom  panel,  the black lines depict power-law fits $P(n) \propto n^{-\beta} $  with $\beta=2.5$.
 }  
\label{fig:7}  
\end{figure}

In the study of casual groups,  a primary quantity of interest is the mean fraction of groups of a given size $n$,  with  $n=1, \ldots, N_{act}$,   at equilibrium \cite{Coleman_1961,White_1962,Fontanari_2023}.  This fraction can be interpreted as the probability of observing a group of size $n$,  denoted as $P(n)$.  Empirical studies typically aggregate observations of groups of the same size across numerous occasions---for instance,  counting pedestrian groups on a sidewalk over many spring mornings  \cite{Coleman_1961}.  They then report the ratio of the total count of groups of a specific size to the total number of groups observed.  In our computational context,  this approach is equivalent to combining groups identified from multiple independent simulation runs.  Figure \ref{fig:7}  summarizes our findings for $P(n)$ across three distinct density values.

For the null model,  $P(n)$ exhibits an exponential decay for large $n$,  provided that  $\rho < \rho_c$.  For densities exceeding $\rho_c$,  the distribution of group sizes becomes bimodal,  though the secondary peak is not  visible at the scale of Fig.\  \ref{fig:7} (top panel).  It features a prominent,  high peak at $n=1$ (isolated agents) and a smoother,  secondary peak for $n$ values close to $N_{act} \approx N$. This behavior is consistent with findings for the RGG \cite{Dall_2002}  and reflects the emergence of a large group that encompasses most of the active agents in each simulation run.   (A full visualization of the null model bimodal distribution and critical scaling is provided in Fig.\ \ref{fig:S3} of the Supplementary Material.) The  average size of this largest group is shown in  Fig.\  \ref{fig:5}.

\begin{figure} 
\centering  
 \includegraphics[width=0.8\textwidth]{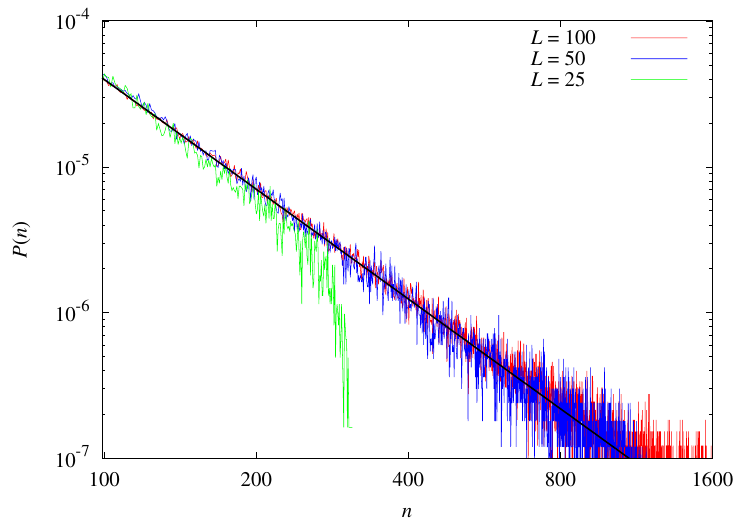}  
 \caption{Group size distribution dependence on system size at $t=10^5$.   Distribution of group sizes $P(n)$  for  the attractiveness-driven interaction model  at $\rho=0.5$.   Results are shown for linear sizes $L=25,  50$,  and $100$.  These  distributions were estimated using  $10^5$  independent simulation  runs.   The black line represents  the  power-law fit $P(n) \propto n^{-\beta} $  with  $\beta=2.5$.  }  
\label{fig:8}  
\end{figure}

Characterizing the group size distribution for the attractiveness-driven interaction model,   shown in the bottom panel of Fig.\  \ref{fig:7},  proves to be quite challenging.  While a power-law decay was initially suggested in a previous study of the model,  the relatively small number of agents considered in that work($N=400$) does not permit such an inference \cite{Starnini_2016}.   Furthermore,  direct fitting of the asymptotic behavior of  $P(n)$  using a variety of fit distributions \cite{Clauset_2009} proved inconclusive,  primarily due to a strong dependence on the chosen fit range.  However,  a valuable insight about the fit range is gained by examining the dependence of $P(n)$ on the number of agents $N$ (or the linear size $L$ of the system) at a fixed density,  as presented in Fig.\  \ref{fig:8}.  
The value $\rho=0.5$  was selected as it allows us to obtain extensive statistics across the three system sizes considered. 

The group size distribution for $L=25$ ($N=312$)  clearly exhibits a sharp cutoff around $n=312$,  which corresponds to the maximum possible group size for that arena.  By extension,  we infer that the cutoffs for $L=50$ and $L=100$ should occur around $n=1250$ and $n=5000$,  respectively.  Our approach for defining the asymptotic fit region is to select the range just before the onset of this cutoff for a given system size.  As depicted in Fig.\ \ref{fig:8},  a power-law $P(n) \propto n^{-\beta} $  with $\beta=2.5$ provides an excellent fit to the data.  However,  a significant challenge arises from the fact that this asymptotic regime sets in for $n>100$,  where $P(n)$ is already very small.  Consequently,  producing reliable statistics necessitates an extremely large number of samples.   For instance,  the distribution for $L=100$  required approximately $10^8$ groups for its estimation.  This makes demonstrating the power-law fit over two or more orders of magnitude practically unfeasible.  The situation further exacerbates with increasing density,  as the asymptotic regime shifts to even larger $n$.  Nevertheless,  having now identified a reliable fit region,  we find that the same power-law also describes the data accurately for high densities (see bottom panel of Fig.\  \ref{fig:7}). This consistency is not surprising,  given our earlier analysis (see  Fig.\  \ref{fig:5}) demonstrating that the attractiveness-driven interaction model's properties are not critically dependent on density.

 \section{Conclusion}\label{sec:disc}
   The attractiveness-driven interaction model introduced by Starnini et al.  \cite{Starnini_2013,Starnini_2016} has significantly advanced our understanding of human contact dynamics, particularly in describing the temporal dimension of face-to-face interactions. While  their original studies primarily focused on distributions of contact times and inter-event times, our research extends this framework by characterizing the sizes of casual,  or temporary,  groups formed by agents within their contact zones. Aligning with the foundational studies of casual groups from the 1960s  \cite{Coleman_1961, White_1962}, our analysis centers on the equilibrium regime, where macroscopic quantities such as average degree and the fraction of agents in the largest group stabilize over time (see Fig.\ \ref{fig:3}).

In addition to introducing a spatial dimension to the casual group formation dynamics, which requires defining groups as connected components of the contact graph,  the attractiveness-driven interaction model  uniquely incorporates the notion of an agent's attractiveness or social appeal $a_i$ for $i=1, \ldots,N$.   This appeal  is associated with the probability that other agents will pause their movement  when in contact with agent $i$ \cite{Starnini_2013}.   We followed the original model by considering $a_i$ to be uniformly distributed within the unit interval.   Agents with high social appeal,  specifically those with $a_i \approx 1$,  can be conceptually considered celebrities. This represents  a realistic feature to be included in models of casual human groups,  a characteristic that previous foundational studies  \cite{White_1962} explicitly omitted. 
We found that the presence of these celebrities profoundly impacts the statistical properties of casual groups, yielding striking differences when compared to the null (agent-homogeneous) model, where $a_i = 0$ for all agents, implying continuous movement for all active participants.

The most significant distinction between the attractiveness-driven interaction model and the null model lies in  the average degree $k$ of the contact graph (see Fig.\ \ref{fig:4}).  In the null model, as in the Random Geometric Graph (RGG), $k$ is solely a function of density.  In sharp contrast,  for  the attractiveness-driven interaction model,  $k$ increases linearly with the arena's linear size $L$ (or with  $N^{1/2}$) for fixed density.  This implies  that as both $L$ and $N$ increase simultaneously to maintain constant density,  the number of agents within the  interaction zone (i.e.,  within a distance of $d=1$) of a given   agent  remains constant in the null model, but increases linearly with $L$ in the attractiveness-driven interaction model.  Consequently,  groups become more compact as $L$ increases.  This clustering of agents into compact groups around celebrities prevents the abrupt formation of a single large group encompassing most agents in the arena.  In fact,  the equilibrium fraction of agents in the largest group ($s_\infty$) increases smoothly with density, notably lacking the percolation transition observed in the null model and the RGG  (see Fig.\ \ref{fig:5}). 

The distribution of group sizes $P(n)$ is arguably the simplest quantitative information  obtainable from  direct observation of casual groups.  For small  gatherings ($N=20$) without celebrity effects,  empirical results are often well described by a zero-truncated Poisson distribution \cite{Coleman_1961}.  
The situation becomes more complex for larger gatherings ($N = 400$),  where empirical distributions have been suggested to be compatible with power-law behavior \cite{Starnini_2016}.   Our extended analysis of the attractiveness-driven interaction model using system sizes up to  $N=5000$  indicates that $P(n)$ is well described by a power law in its asymptotic regime,  specifically  $P(n) \propto n^{-\beta}$.   Although computationally very challenging to reach and reliably characterize this regime,  the power-law exponent  appears to be independent  of density, with a value of  $\beta = 2.5$ (see Figs.\ \ref{fig:7}  and  \ref{fig:8}).

The most profound structural implication of the attractiveness mechanism is the suppression of the percolation phase transition, a ubiquitous phenomenon in proximity networks such as the Random Geometric Graph \cite{Dall_2002}. This suppression is a robust conclusion supported by the consistency across our key equilibrium observables. The scaling of the average degree $k_\infty \propto L$  demonstrates the physical mechanism: the intensification of local, compact clusters around celebrities as the system size grows. This highly heterogeneous spatial organization effectively traps agents in localized groups, preventing the formation of a macroscopic spanning cluster. 
Consequently, the fraction of agents in the largest group,  $s_\infty$, increases smoothly with density $\rho$ (see Fig.\  \ref{fig:5}), maintaining a continuous and gradually increasing profile across the density range where the  null model exhibits an abrupt jump and a bimodal group size distribution (see Fig.\  \ref{fig:S3}, Supplementary Material). This ability of the attractiveness-driven interaction model to self-organize into dense, stable local clusters while avoiding global connectivity signifies a fundamental shift in the universal behavior of spatial interaction models driven by agent heterogeneity.

This study highlights the critical role of individual social appeal in shaping the macro-scale structure and dynamics of human contact networks,  effectively bridging contemporary agent-based modeling  \cite{Starnini_2013,Starnini_2016} with classical sociological insights into casual group formation  \cite{Coleman_1961, White_1962}.  By demonstrating how the heterogeneous distribution of attractiveness,  or the presence of celebrities,  fundamentally alters group cohesion, connectivity, and emergent properties such as percolation,  our re-examination of  the attractiveness-driven interaction model   offers a thorough quantitative characterization.   Future work will explore the implications of attractiveness distributions that more closely mirror real-world social hierarchies, such as wealth distribution  \cite{Newman_2005},  and consider the inclusion of a repulsive hard core for  agents,  mimicking the comfort distance observed in humans \cite{Dosey_1969} and animals \cite{Okubo_1986},  thereby further enriching the predictive power of these models for social aggregation in physical space.  Additionally, new generalizations of the model that incorporate higher-order interactions offer exciting avenues for future research \cite{Gallo_2024}.

\backmatter

\bmhead{Acknowledgments}
JFF is partially supported by  Conselho Nacional de Desenvolvimento Cient\'{\i}fico e Tecnol\'ogico  grant number 305620/2021-5.  
MSM is  supported by Fun\-da\-\c{c}\~ao de Amparo \`a Pesquisa do Estado de S\~ao Paulo 
(FAPESP) grant number  2024/00582-4.
This research is partially supported by Funda\c{c}\~ao de Amparo \`a Ci\^encia e Tecnologia do Estado de Pernambuco (FACEPE) grant number APQ-1129-1.05/24.





\newpage

\begin{center}
\section*{Supplementary Material}
\end{center}

 \setcounter{figure}{0}
 \renewcommand{\thefigure}{S\arabic{figure}}

This Supplementary Material provides additional quantitative evidence, technical justifications, and detailed characterizations that support the central findings presented in the main manuscript. The contents directly address specific points raised during the peer review process, offering deeper insight into: (1) the time scales required for the system to reach equilibrium, (2) the scaling behavior of the active population, and (3) a comprehensive analysis of the equilibrium spatial and network structure, including a detailed comparison with the null model.

\bigskip

\section*{\normalsize{Dependence of the relaxation time on system size}}\label{sec:SM1}

\begin{figure} [h]
\centering  
 \includegraphics[width=0.8\textwidth]{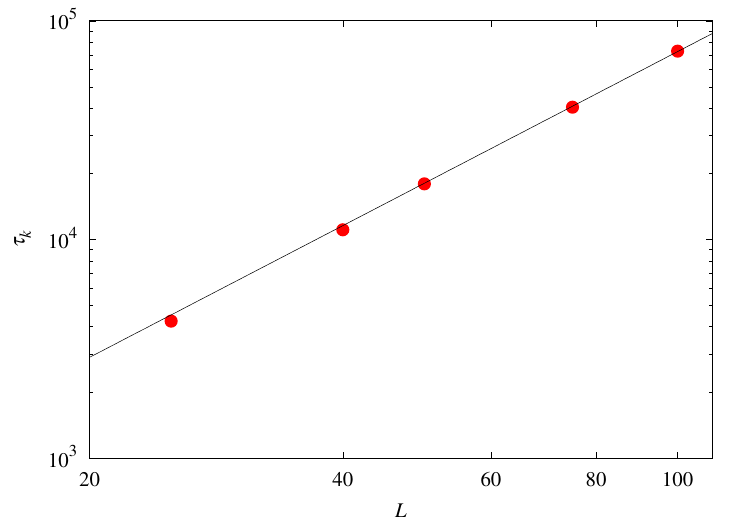}  
 \caption{Relaxation time $\tau_k$ of the average degree $k$ as a function of the system size $L$ for the attractiveness-driven interaction model at fixed density $\rho=0.6$. Data points (symbols) are averages over $10^3$ independent simulation runs. The solid line represents the best fit to the power-law ansatz $\tau_k = a L^b$, yielding $a \approx 7.24$ and the exponent $b = 2.00$ which precisely confirms the expected $\tau_k \propto L^2$ scaling.
  }  
\label{fig:S1}  
\end{figure}

In the main text (right panel of Fig.\ \ref{fig:3}), we fit the time evolution of the average degree $k (t)$ using eq.  (\ref{fitk}) and observe that the characteristic relaxation time $\tau_k$ increases with increasing system size $L$. To provide a quantitative justification for this observation, Fig.\ \ref{fig:S1} shows the dependence of $\tau_k$ on $L$ at the fixed density $\rho=0.6$ used in the main text. The data are excellently described by the ansatz $\tau_k \propto L^2$, confirming that the relaxation time scales quadratically with the linear dimension. This scaling establishes the total number of agents  $N=\rho L^2$ as the primary limiting factor for simulating large systems, necessitating the long transient periods discussed in the paper.
 
\section*{\normalsize{Scaling of the active population}}\label{sec:SM2}

\begin{figure}[h] 
\centering  
 \includegraphics[width=0.8\textwidth]{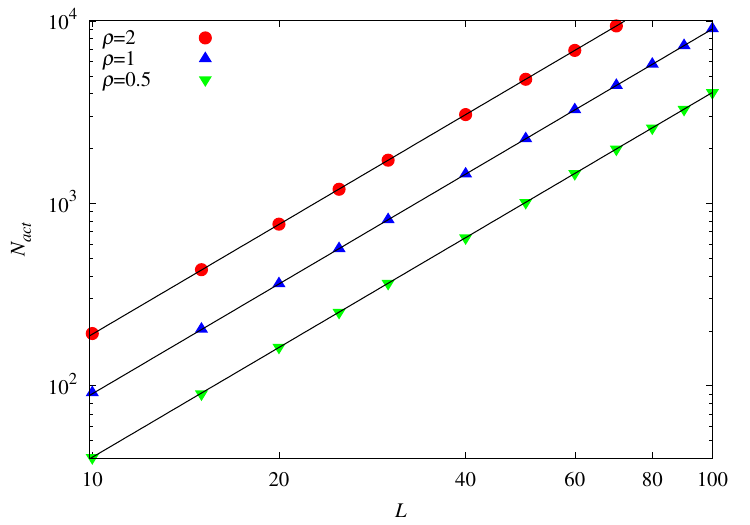}  
 \caption{Log-log plot of the average number of active agents $N_{act} $ as a function of the system size $L$ for the attractiveness-driven interaction model at fixed densities $\rho=0.5, 1$, and $2$.  Data points (symbols) represent averages obtained from  $10^3$  independent simulation runs. The solid lines represent the best fits to the ansatz $N_{act} = a L^2$. The resulting coefficients $a$  are approximately $0.405$  ($\rho=0.5$),  $0.903$  ($\rho=1$), and $1.916$  ($\rho=2$). 
}  
\label{fig:S2}  
\end{figure}

The  average number of active agents, $N_{act}$, is a key dynamic variable of the attractiveness-driven interaction model. In the main text, we assume  that $ N_{act} $ scales linearly with the total number of agents $N$ for fixed density $\rho$ or, equivalently,  that  $ N_{act} \propto L^2 $. This scaling is quantitatively  confirmed in Fig.\  \ref{fig:S2}, which presents  $N_{act} $ as a function of the linear system size $L$ for  $\rho=0.5, 1 $, and $2$.  The straight lines with slope $2$ in the log-log plot directly validate this quadratic scaling  under fixed-density conditions.  The coefficient  $a$ in the ansatz $N_{act} = a L^2$ is  less than $\rho$ since  $N_{act} < N$. As $\rho \to 0$ (low density), most agents are isolated, and $ N_{act}  \approx \rho L^2/2$, given that the average inactivation probability for isolated agents is  $1/2$.  Conversely, as $\rho \to \infty$ (high density), virtually no agent is isolated, leading to $ N_{act}  \approx  \rho L^2$.

\section*{\normalsize{Cluster size distribution of the null model}}\label{sec:SM3}

The analysis in the main text compares the behavior of the attractiveness-driven interaction model  to the null model,   which for large density is equivalent to a Random Geometric Graph (RGG) at equilibrium. As shown in Fig. \ \ref{fig:5}, the RGG undergoes a standard percolation phase transition at a critical density $\rho_c \approx 1.44$. The cluster size distribution, $P(n)$, exhibits a characteristic power-law dependence on the group size $n$  near this critical point \cite{Dall_2002}.

\begin{figure} 
\centering  
 \includegraphics[width=0.8\textwidth]{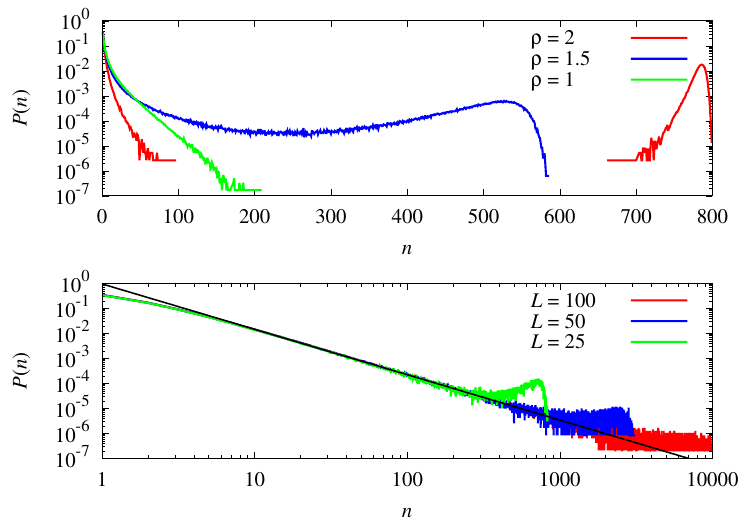}  
 \caption{Cluster size distribution, $P(n)$, for the null model (Random Geometric Graph).  The upper panel shows $P(n)$ for fixed system size $L=20$ at densities below ($\rho=1$),  slightly above ($\rho=1.5$), and well above ($\rho=2$) the critical density  $\rho_c \approx 1.44$.   The distributions for $\rho \geq \rho_c$ clearly show the bimodal nature, with a second peak at large $n$ indicating the emergence of the spanning cluster.
 The lower panel shows $P(n)$ at the critical density $\rho_c  $ for different system sizes $L=25, 50$, and $100$.  The solid curve is the power-law fit $P(n) =a  n^{-\tau}$ with $a=0.95$ and $\tau = 1.81$.}  
\label{fig:S3}  
\end{figure}

Figure \ref{fig:S3} confirms the known behavior of the RGG.  The upper panel explicitly demonstrates the emergence of bimodality in $P(n)$ when the density crosses $\rho_c$. For $\rho=1.5$ and $\rho=2$  (where $\rho > \rho_c$), the cluster size distribution shows a second peak at large group sizes $n \approx N$, corresponding to the formation of a single spanning cluster.  This bimodal feature is completely absent  in the attractiveness-driven interaction  model and provides the key context for understanding the suppressed percolation transition observed in our main model.  The lower panel shows the power-law decay of $P(n) \propto n^{-\tau}$ with $\tau \approx 1.81$ at the critical density $\rho_c \approx 1.44$ for large system sizes $L=25, 50$, and $100$.

\section*{\normalsize{The effect of density on the pair correlation function}}\label{sec:SM4}

\begin{figure} 
\centering  
 \includegraphics[width=0.8\textwidth]{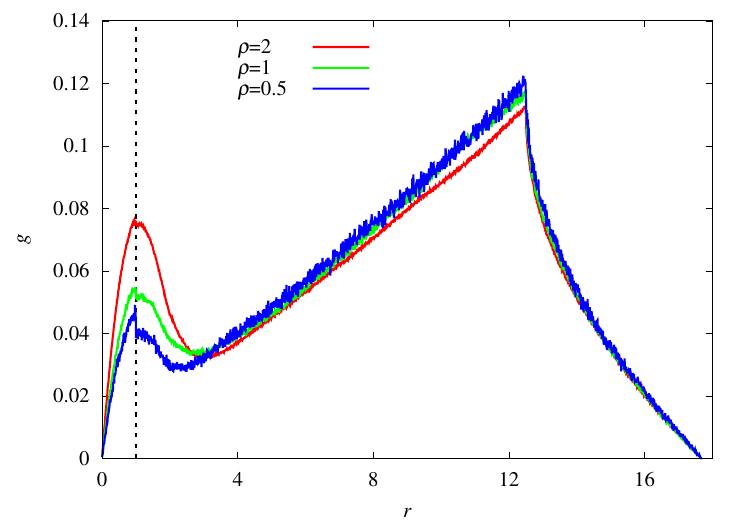}  
 \caption{Pair correlation function $g(r)$ at equilibrium for the attractiveness-driven interaction model at densities $\rho=0.5, 1$, and $2$. The system size is $L=25$, and results are averaged over 100 independent simulation runs after the system reached its steady state ($t=2 \times 10^4$). The vertical dashed line indicates the interaction distance $r=d=1$.}  
\label{fig:S4}  
\end{figure}

The finding that increasing the density $\rho$ makes the clusters more compact around the celebrities is crucial for the  observed  suppression of the percolation transition.  This effect is  directly  reflected in the pair correlation function $g(r)$,   which is shown in Fig.\  \ref{fig:S4} for $L=25$ and several values of $\rho$.  The figure clearly demonstrates how clustering intensifies with increasing  $\rho$,  as $ g(r)$ develops  a significantly higher peak at the interaction distance $r=d=1$.  This elevated peak signifies an increased probability of finding an agent at the precise distance where interaction occurs, confirming that agents aggregate into denser local environments. Importantly, the peak remains broad, indicating that while local density increases, the characteristic size of the local cluster does not fundamentally change across this range of densities.

Note that the decrease in statistical fluctuations visible in $g(r)$ with increasing $\rho$ (or, equivalently, total number of agents $N$, since $L$ is fixed) is a direct consequence of the increasing  sample size. As discussed in the main text, the evaluation of $g(r)$ uses the number of unique pairs, $N(N-1)/2$, which increases quadratically with $N$, naturally leading to smoother curves at higher densities.

\section*{\normalsize{The equilibrium degree distribution}}\label{sec:SM5}

\begin{figure} 
\centering  
 \includegraphics[width=0.8\textwidth]{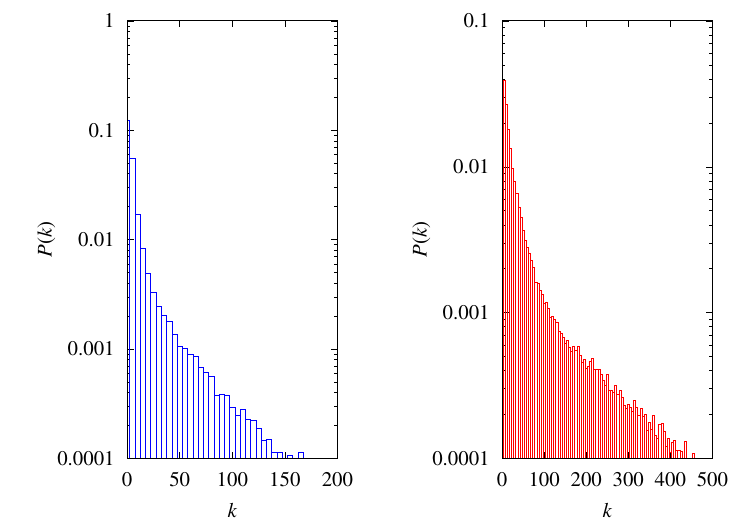}  
 \caption{Histogram of the  degree distribution, $P(k)$, at equilibrium for the attractiveness-driven interaction model at densities $\rho=0.5$ (left panel) and $\rho=2$ (right panel). The system size is $L=25$, and results are averaged over 1000 independent simulation runs after the system reached its steady state ($t=2 \times 10^4$).}  
\label{fig:S5}  
\end{figure}

In the main text, we focused only on the equilibrium average degree $ k_{\infty}$  of the face-to-face interaction network. To provide a more comprehensive characterization of the network's local structure, Fig.\  \ref{fig:S5} displays the  degree distribution, $P(k)$, for the attractiveness-driven interaction model at equilibrium for two different densities.

The distributions are broad and decay smoothly, which confirms that the presence of high-attractiveness agents (celebrities) does not  induce multimodal distributions in the interaction network, thereby addressing a potential concern about the  network structure. The single, broad peak shifts toward larger $k$ as density increases, consistent with the linear increase in $ k_{\infty}$ with $\rho$ shown in Fig. \ \ref{fig:4} of the main text. For large $k$, we observe that $P(k)$ vanishes rapidly, consistent with an exponential decay with increasing degree.

\end{document}